\documentclass{article}

\begin{document}

\title{On the gravitational redshift}
\author{Vesselin Petkov \\
Physics Department, Concordia University\\
1455 De Maisonneuve Boulevard West\\
Montreal, Quebec, Canada H3G 1M8\\
E-mail: vpetkov@alcor.concordia.ca}
\date{3 September 2001}
\maketitle

\begin{abstract}
The purpose of this paper is twofold - to demonstrate that in the
gravitational redshift it is the frequency a photon that is
constant, and to describe the mechanism responsible for the change
of its wavelength.

\medskip

\end{abstract}

It is usually assumed that both frequency and wavelength of a
photon in the gravitational redshift change whereas its velocity
remains constant. In this note we shall show that it is the
frequency of a photon that does not change whereas its velocity
and wavelength change. It will be also shown that it is the change
of the coordinate velocity of the photon along its path that leads
to a change in its wavelength.

Three things should be kept in mind when dealing with the gravitational
redshift:

1. If two observers at different points $A$ and $B$ in a gravitational field
determine the characteristics of a photon emitted from identical atoms
placed at $A$ and $B$, each observer will find that the photon
characteristics - frequency, wavelength and local velocity - will have the
same numerical values.

2. In a \emph{parallel} gravitational field coordinate and proper distances
coincide $dx=dx_{A}=dx_{B}$ \cite{rindler68} and therefore the wavelength of
a photon \emph{at a point} is the same for all observers - $\lambda
_{A}=\lambda _{B}=\lambda $.

3. The local velocity of a photon \emph{at a point} is different for
different observers (it is $c$ only for an observer at that point).

Consider a non-inertial frame $N^{g}$ at rest in a \emph{parallel}
gravitational field of strength $\mathbf{g}$. If the $z$-axis is
anti-parallel to the acceleration $\mathbf{g}$ the spacetime metric in $N^{g}
$ has the form \cite{misner}
\begin{equation}
ds^{2}=\left( 1+\frac{2gz}{c^{2}}\right) c^{2}dt^{2}-dx^{2}-dy^{2}-dz^{2}
\label{ds_g}
\end{equation}
from where the coordinate velocity of light at a point $z$ in a parallel
gravitational field is immediately obtained (for $ds^{2}=0$)
\begin{equation}
c^{g}=c\left( 1+\frac{gz}{c^{2}}\right) .  \label{c_g_coord}
\end{equation}
Notice that (\ref{ds_g}) is the standard spacetime interval in a \emph{%
parallel} gravitational field \cite{misner}, which does not coincide with
the expression for the spacetime interval in a spherically symmetric
gravitational field (i.e. the Schwarzschild metric expressed here in
Cartesian coordinates) \cite[p. 395]{ohanian}
\begin{equation}
ds^{2}=\left( 1-\frac{2GM}{c^{2}r}\right) c^{2}dt^{2}-\left( 1+\frac{2GM}{%
c^{2}r}\right) \left( dx^{2}+dy^{2}+dz^{2}\right) .  \label{schwarz}
\end{equation}
The metric (\ref{ds_g}) can be written in a form similar to (\ref{schwarz})
if we choose $r=r_{0}+z$ where $r_{0}$ is a constant
\begin{equation}
ds^{2}=\left( 1-\frac{2GM}{c^{2}\left( r_{0}+z\right) }\right)
c^{2}dt^{2}-\left( dx^{2}+dy^{2}+dz^{2}\right) .  \label{ds_g1}
\end{equation}
As $g=GM/r_{0}^{2}$ and for $z/$ $r_{0}<1$ we can write
\begin{equation}
ds^{2}=\left( 1-\frac{2GM}{c^{2}r_{0}}+\frac{2gz}{c^{2}}\right)
c^{2}dt^{2}-\left( dx^{2}+dy^{2}+dz^{2}\right) .  \label{ds_g2}
\end{equation}
As the gravitational potential is undetermined to within an additive
constant we can choose $GM/r_{0}=0$ in (\ref{ds_g2}); more precisely, when
calculating the gravitational potential we can set the constant of
integration to be equal to $-GM/r_{0}.$ With this choice of the integration
constant (\ref{ds_g2}) coincides with (\ref{ds_g}). Although similar (\ref
{ds_g1}) and (\ref{schwarz}) have different values for $g_{ii}$ $(i=1,2,3)$:
$g_{ii}=-1$ in (\ref{ds_g1}), whereas $g_{ii}=-\left( 1+2GM/c^{2}r\right) $
in (\ref{schwarz}). This reflects the fact that in a parallel gravitational
field proper and coordinate times do not coincide (except for the proper
time of an observer at infinity) whereas proper and coordinate distances are
the same \cite{rindler68}.

Consider an atom stationary at a point~$B$ in a parallel gravitational
field. The atom emits a photon - a $B$-photon - which is observed at a point~%
$A$ at a distance $h$ above $B$. As seen at $B$ the $B$-photon is emitted
with a frequency $f_{B}=\left( d\tau _{B}\right) ^{-1}$, where $d\tau _{B}$
is the proper period. As seen from $A$, however, the $B$-photon's period is $%
d\tau _{A}$ and therefore its frequency is $f_{B}^{A}=\left( d\tau
_{A}\right) ^{-1}$. Notice that if an identical atom at $A$ emits a photon
its frequency at $A$ will be $f_{A}=\left( d\tau _{A}\right) ^{-1}=f_{B}$,
which means that the corresponding periods will be (numerically) equal: $%
d\tau _{A}=d\tau _{B}$. In the case of the redshift experiment, however,
when a $B$-photon is measured at $A$, $d\tau _{A}$ and $d\tau _{B}$ are
different - $d\tau _{B}$ is the proper period (measured at $B$) whereas $%
d\tau _{A}$ is the \emph{observed} period as measured at $A$. $d\tau _{A}$
and $d\tau _{B}$ are the proper times at $A$ and $B$ that correspond to the
\emph{same} coordinate time, i.e. the same coordinate period $dt$:
\[
d\tau _{A}=\left( 1+\frac{gz_{A}}{c^{2}}\right) dt
\]
and
\[
d\tau _{B}=\left( 1+\frac{gz_{B}}{c^{2}}\right) dt.
\]
As $z_{A}=z_{B}+h$ it follows from (\ref{ds_g}) that the ratio between $%
d\tau _{A}$ and $d\tau _{B}$ is
\[
\frac{d\tau _{A}}{d\tau _{B}}=\frac{\left( 1+gz_{A}/c^{2}\right) }{\left(
1+gz_{B}/c^{2}\right) }\approx 1+\frac{gh}{c^{2}}.
\]
Therefore, the \emph{initial} frequency of the $B$-photon at $B$\ as seen
from $A$ will be
\begin{equation}
f_{A}=\frac{1}{d\tau _{A}}=\frac{1}{d\tau _{B}\left( 1+gh/c^{2}\right) }%
\approx f_{B}\left( 1-\frac{gh}{c^{2}}\right) .  \label{f_A}
\end{equation}
As seen from (\ref{f_A}) for an observer at $A$ the $B$-photon is emitted
with a reduced initial frequency $f_{A}<f_{B}$. This demonstrates that the
frequency of the $B$-photon does not change during its journey from $A$ to $B
$ since its final frequency at $A$ should be also (\ref{f_A}).

The same expression for the initial frequency of the $B$-photon at $B$\ as
seen from $A$ can be obtained if one makes use of the fact that in a
parallel gravitational field proper and coordinate distance coincide. This
means that the initial wavelength $\lambda _{A}$ of the $B$-photon at $B$\
as seen from $A$ is equal to the initial wavelength $\lambda _{B}$ as
measured at $B$ - $\lambda _{A}=\lambda _{B}=\lambda $. The initial velocity
of the $B$-photon at $B$\ as seen from $A$ can be easily calculated
\[
c_{A}=\frac{dz_{B}}{d\tau _{A}}=\frac{dz_{B}}{dt}\frac{dt}{d\tau _{A}}
\]
where and $dz_{B}/dt$ is the coordinate velocity at point $B$

\[
c^{\prime }=c\left( 1+\frac{gz_{B}}{c^{2}}\right)
\]
and

\[
dt=\left( 1-\frac{gz_{A}}{c^{2}}\right) d\tau _{A}.
\]
As $z_{A}=z_{B}+h$ we can write

\begin{equation}
c_{A}=c\left( 1-\frac{gh}{c^{2}}\right)  \label{c_A}
\end{equation}

Hence, the frequency of the $B$-photon at $B$\ as seen from $A$ is
\[
f_{A}=\frac{c_{A}}{\lambda }=f_{B}\left( 1-\frac{gh}{c^{2}}\right)
\]
where $f_{B}=c/\lambda $.

The fact that the $B$-photon's frequency does not change demonstrate that
its energy is constant - an indication that the photon is not losing energy
while moving against the gravitational field. Inversely, if an $A$-photon is
observed at $B$\ its constant energy will indicate that it is not gaining
energy and therefore is not falling in the gravitational field (if it were
falling its average downward speed would be greater than its upward average
speed which is not the case ).

We have seen that it is the frequency that is constant - a conclusion also
pointed out by Okun, Selivanov, and Telegdi \cite{okun}. What changes as the
$B$-photon travels toward the observation point $A$ , as seen from $A$, is
its velocity and wavelength. The initial velocity of the $B$-photon at $B$ ,
as seen from $A$, is given by (\ref{c_A}); its final velocity at $A$ , as
seen from $A$, should be obviously $c$. The change of the photon's velocity
on its way toward $A$ also explains the mechanism responsible for the change
of its wavelength. As seen from $A$ any wavefront moving away from the
gravitational field (toward $A$) acquires a greater velocity as compared to
the velocity of the next wavefront that follows it. Due to the speeding up
of the first wavefront the spacing between the two wavefronts increases for
one period $d\tau _{A}$ (as seen by $A$) by a fraction $\delta \lambda
=\delta c\ d\tau _{A}$ where
\[
\delta c=c\left[ 1+\frac{g\left( z+dz\right) }{c^{2}}\right] -c\left( 1+%
\frac{gz}{c^{2}}\right) =c\frac{gdz}{c^{2}}
\]
is the change of the coordinate velocity over the distance $dz$.

Then the total increase of the wavelength from $B$ to $A$ is

\[
\Delta \lambda =\int_{0}^{h}\delta c\ d\tau _{A}=c\frac{gd\tau _{A}}{c^{2}}%
\int_{0}^{h}dz=c\frac{gh}{c^{2}}d\tau _{A}.
\]
As

\[
d\tau _{A}=d\tau _{B}\left( 1+\frac{gh}{c^{2}}\right)
\]
we can write for $\Delta \lambda $ by keeping only the terms proportional to
$c^{-2}$

\[
\Delta \lambda =c\frac{gh}{c^{2}}d\tau _{B}=\lambda \frac{gh}{c^{2}}
\]
where $c\ d\tau _{B}=\lambda $ is the initial wavelength as determined at $B$%
. The final (measured) wavelength of the $B$-photon at $A$ is then

\[
\lambda _{A}=\lambda +\Delta \lambda =\lambda \left( 1+\frac{gh}{c^{2}}%
\right) .
\]

Therefore, in the gravitational redshift it is the velocity and wavelength
of a photon that change whereas its frequency does not change.

\end{document}